\documentclass[useAMS]{mn2e}
\usepackage{epsfig}
\usepackage{cases}
\usepackage{graphicx}
\usepackage{latexsym}

\def\lsim{\,\lower2truept\hbox{${<\atop\hbox{\raise4truept\hbox{$\sim$}}}$}\,}
\def\gsim{\,\lower2truept\hbox{${> \atop\hbox{\raise4truept\hbox{$\sim$}}}$}\,}

\title[Clustering of dusty galaxies]{Clustering of sub-millimeter galaxies in a self-regulated baryon collapse model}
\author[J.-Q. Xia et al.]{
\parbox[t]{\textwidth}
{Jun-Qing Xia$^{1}$\thanks{E-mail: xia@sissa.it}, M. Negrello$^{2}$, A. Lapi$^{3,1}$, G. De Zotti$^{4,1}$, L. Danese$^{1}$, M. Viel$^{5,6}$}\\
\vspace*{8pt} \\
$^{1}$Scuola Internazionale Superiore di Studi Avanzati, via Bonomea 265, I-34136 Trieste, Italy\\
$^{2}$Dept. of Physics \& Astronomy, The Open Univ., Milton Keynes MK7 6AA, UK\\
$^{3}$Dip. Fisica, Univ. `Tor Vergata', Via Ricerca Scientifica 1, 00133 Roma, Italy\\
$^{4}$INAF-Osservatorio Astronomico di Padova, vicolo dell'Osservatorio 5, I--35122 Padova, Italy\\
$^{5}$INAF-Osservatorio Astronomico di Trieste, Via G.B. Tiepolo 11, I-34131 Trieste, Italy\\
$^{6}$INFN sez. Trieste, Via Valerio 2, 34127 Trieste, Italy}

\begin{document}

\date{}
%
\pagerange{1--8} \pubyear{2011}
\maketitle
\label{firstpage}

\begin{abstract}
We have investigated the Cosmic Infrared Background (CIB)
anisotropies in the framework of the physical evolutionary model for
proto-spheroidal galaxies by Granato et al. (2004). After having
re-calibrated the cumulative flux function $dS/dz$ at $\lambda \ge
850\,\mu$m using the available determinations of the shot noise
amplitude (the original model already correctly reproduces it at
shorter wavelengths) the CIB power spectra at wavelengths from
$250\,\mu$m to $2\,$mm measured by {\it Planck}, {\it Herschel}, SPT
and ACT experiments have been fitted using the halo model with only
2 free parameters, the minimum halo mass and the power-law index of
the mean occupation function of satellite galaxies. The best-fit
{\it minimum} halo mass is $\log(M_{\rm min}/M_\odot) = 12.24 \pm
0.06$, higher than, but consistent within the errors, with the
estimate by Amblard et al. (2011) and close to the estimate by
Planck Collaboration (2011). The redshift evolution of the volume
emissivity of galaxies yielded by the model is found to be
consistent with that inferred from the data. The derived {\it
effective} halo mass, $M_{\rm eff} \simeq 5\times 10^{12}\,M_\odot$,
of $z\simeq 2$ sub-millimeter galaxies is close to that estimated
for the most efficient star-formers at the same redshift. The
effective bias factor and the comoving clustering radius at $z\simeq
2$ yielded by the model are substantially lower than those found for
a model whereby the star formation is fueled by steady gas
accretion, but substantially higher than those found for a
merging-driven galaxy evolution with a top-heavy initial mass
function.
\end{abstract}

\section{Introduction} \label{sec:intro}
The {\it Herschel}  surveys have allowed clustering studies (Maddox
et al. 2010; Cooray et al. 2010) of sub-millimeter galaxies with a
statistics at least one order of magnitude better than previously
possible (Blain et al. 2004; Scott et al. 2006). These studies have
been complemented by determinations of the angular power spectrum of
the Cosmic Infrared Background (CIB) anisotropies on BLAST (Viero et
al. 2009), {\it Planck} (Planck Collaboration 2011), and {\it
Herschel}  (Amblard et al. 2011) maps. Due to the unique power of
sub-millimeter surveys in piercing the distant universe, thanks to
the strongly negative K-correction, the clustering properties
contain signatures of the large scale structure at high redshifts
and can allow us to discriminate between different formation
mechanisms that have been proposed for sub-millimeter galaxies. For
example, merger driven galaxy evolution models, that follow the
evolution of both the disk and the spheroidal components of
galaxies, predict much lower clustering strengths for sub-mm
galaxies (e.g. Almeida et al. 2011; Kim et al. 2011) than models
whereby the star formation is fueled by steady accretion of large
amounts of cold gas (e.g. Dav\'e et al. 2010).

In this paper, building on the work by Negrello et al. (2007), we
investigate the constraints set by mm and sub-mm clustering data on
the physical model worked out by Granato et al. (2001, 2004) and
further elaborated by Lapi et al. (2006) and Mao et al. (2007).

A specific prediction of the model is that high-$z$ massive
proto-spheroidal galaxies dominate the sub-mm counts over a limited
flux density range (cf. Lapi et al. 2011). At $250\,\mu$m the
Euclidean normalized differential counts of these objects peak at
$\approx 30$ mJy; above $\simeq 60\,$mJy and below $\simeq 10\,$mJy
the counts are dominated by $z\lsim 1.5$ quiescent and star-bursting
late-type galaxies, less massive and less clustered than the
high-$z$ proto-spheroidal galaxies. The flux density range where
proto-spheroidal galaxies dominate broadens and the peak shifts to
brighter flux densities with increasing (sub-)mm wavelength.
Therefore, in this scenario, the expected clustering strengths
depend on the flux density range that is being probed and on
wavelength.

Several other analyses of data on the angular correlation function
of (sub-)mm sources and of the power spectrum of the CIB
anisotropies have been carried out. They however use
phenomenological parameterized models for the evolution of
extragalactic sources (Hall et al. 2010; Planck Collaboration 2011;
Millea et al. 2011; P\'enin et al. 2011) or even of the clustering
power (Addison et al. 2011). Also data at different wavelengths are
usually fitted separately (Planck Collaboration 2011; Amblard et al.
2011). On the contrary, the present analysis relies on a physical
model for the evolution of proto-spheroidal galaxies (although the
treatment of spiral and starburst galaxies is phenomenological) and
aims at accounting simultaneously for clustering data over a broad
range of wavelengths, from $250\,\mu$m to a few mm.

It should be noted, however, that the physical model is exploited
only to compute the cumulative flux function that weights the
redshift-dependent spatial power spectrum in the Limber
approximation for the angular power spectrum. The Halo Occupation
Distribution (HOD), which is a statistical description of how dark
matter halos are populated with galaxies, is dealt with in a
simplified manner, without including the relationship between
luminosity and halo mass. This is the standard practice, justified
by the complexity of a thorough treatment that does not appear to be
required by existing data. In the Granato et al. (2004) model the
star-formation rate is related to the halo mass, {\it to the
virialization redshift and to the age of the galaxy}. Including
these additional ingredients in the analysis is impractical at the
present stage. A pioneering model that explicitly includes a
relationship between infrared luminosity and halo mass has been
presented by Shang et al. (2011).

The plan of the paper is the following. In \S\,\ref{sec:model} we
present a short overview of the evolutionary model for the relevant
galaxy populations. In \S\,\ref{sect:hm} we describe the halo model
formalism used to compute the contributions to the power spectrum of
Cosmic Infrared Background (CIB) anisotropies and to the angular
correlation function of detected galaxies (\S\,\ref{sect:ps}). Our
main results are presented in \S\,\ref{sect:results} and our main
conclusions are summarized in \S\,\ref{sect:conclusions}.

We adopt a standard flat $\Lambda$CDM cosmology with
$h=H_0/100\,\hbox{km}\,\hbox{s}^{-1}\,\hbox{Mpc}^{-1}=0.70$ and a
local matter density $\Omega_{{\rm m}0}=0.27$.

\section{Overview of the model} \label{sec:model}

The sub-millimeter extragalactic sources are a mixed bag of various
populations of dusty galaxies and of flat-spectrum radio sources
(see, e.g., Lapi et al. 2011).

Our model interprets powerful high-$z$ sub-mm galaxies as massive
proto-spheroidal galaxies in the process of forming most of their
stellar mass (see also Blain et al. 2004; Narayanan et al. 2010;
Dav\'e et al. 2010). It hinges upon high resolution numerical
simulations showing that dark matter halos form in two stages (Zhao
et al. 2003; Wang et al. 2011; Lapi \& Cavaliere 2011). An early
fast collapse of the halo bulk, including a few major merger events,
reshuffles the gravitational potential and causes the dark matter
and the stellar component to undergo (incomplete) dynamical
relaxation. A slow growth of the halo outskirts in the form of many
minor mergers and diffuse accretion follows; this second stage has
little effect on the inner potential well where the visible galaxy
resides.

The star formation is triggered by the fast collapse/merger phase of
the halo and is controlled by self-regulated baryonic processes. It
is driven by the rapid cooling of the gas within a region of $\simeq
70(M_h/10^{13}\,M_\odot)^{1/3}[(1+z)/3]^{-1}\,$kpc, where $M_h$ is
the halo mass, is regulated by the energy feedback from supernovae
(SNe) and Active Galactic Nuclei (AGNs), is very soon obscured by
dust and is stopped by quasar feedback.  The AGN feedback is
relevant especially in the most massive galaxies and is responsible
for their shorter duration ($5-7\times 10^8\,$yr) of the active
star-forming phase. In less massive proto-spheroidal galaxies the
star formation rate is mostly regulated by SN feedback and continues
for a few Gyr.

Since spheroidal galaxies are observed to be
in passive evolution at $z\la 1-1.5$ (e.g., Renzini 2006), they are
visible at sub-mm wavelength only at high redshifts. Lapi et al.
(2011) have shown that the Granato et al. (2004) model, as further
elaborated by Lapi et al. (2006), provides a reasonably good fit to
the observed counts from $250\,\mu$m to $\simeq 1\,$mm as well as to
the luminosity functions in the range $z=1-4$ and to the redshift
distributions at $z>1$ estimated from {\it Herschel}-ATLAS (Eales et
al. 2010) data.

The fit was obtained using of a single SED (that of the well studied
$z=2.3$ strongly lensed galaxy SMM~J2135-0102, ``The Cosmic
Eyelash''; Ivison et al. 2010, Swinbank et al. 2010) for the whole
population of proto-spheroidal galaxies. This is obviously an
oversimplification and indeed the Lapi et al. (2011) counts are
somewhat high at mm wavelengths, especially at relatively bright
flux densities. As a consequence, the model overestimates the
Poisson (shot-noise) contribution to the power spectrum of intensity
fluctuations since such contribution is directly related to the
source counts [see eq.~(\ref{eq:shot})]. Consistency with the
shot-noise levels estimated by Planck Collaboration (2011) at 353
GHz and measured by Hall et al. (2010), Dunkley et al. (2011) at 220
and 150 GHz is recovered scaling down the cumulative flux function
$dS/dz$ [see eq.~(\ref{eq:dSdz})] of proto-spheroidal galaxies by a
factor of 0.81, 0.71, and 0.55 at 353, 217, and 150 GHz
($850\,\mu$m, 1.38\,mm, 2\,mm), respectively. No correction was
applied at higher frequencies. In practice, we use the determination
of the shot noise amplitude to recalibrate the function $dS/dz$ to
be used to compute the clustering power spectrum, which is measured
independently. This correction mimics the effect of adopting a SED
decreasing with increasing wavelength beyond the peak a bit more
steeply than the one adopted by Lapi et al. (2011).

As suggested in the latter paper, the overestimate of mm-wave counts
may be cured if higher-$z$ galaxies, that yield larger and larger
contributions to the bright counts at increasing mm wavelengths,
have SEDs slightly hotter than SMM J2135-0102 and closer to that of
G15.141 (Cox et al. 2011; see Fig. 2 of Lapi et al. 2011). We have
checked that indeed a good fit of the counts at all the frequencies
considered here is obtained using the SMM J2135-0102 SED for
galaxies at $z<3.5$ and the SED of G15.141 at higher z. However the
match of the frequency spectrum of the shot-noise amplitude also
improves but not enough to reach consistency with observational
estimates at the longer wavelengths. Since the shot noise amplitude
can be computed directly from the counts, this suggests that there
may be some small, but non-negligible, offsets between the
calibration of point source flux densities and that of the diffuse
background. This is not surprising since, in addition to the
possibility of an imperfect photometric calibration, at mm
wavelengths the recovery of the contribution of dusty galaxies to
the power spectrum requires a delicate subtraction of the other
components (Cosmic Microwave Background, cirrus emission,
fluctuations due to radio sources). A rescaling to match the shot
noise spectrum seems to be the only practical way for correcting for
these offsets. Since the modification is only significant at $\ge
850\,\mu$m (in the observer frame), i.e. well beyond the peak for
most sources, the impact on the bolometric luminosity, which is
related to the halo mass, is minor. For galaxies at redshifts up to
$z=3.5$, accounting for essentially all the signal, the bolometric
luminosity varies by $\le 2\%$. For comparison, the coefficient of
the relationship between the star formation rate (SFR; given by the
model) and the bolometric luminosity has an uncertainty of $\sim
30\%$ (Kennicutt 1998).


The Granato et al. (2004) model is meant to take into account the
star formation occurring within galactic dark-matter halos
virialized at $z_{\rm vir} \gsim 1.5$ and bigger than $M_{\rm vir}
\simeq 10^{11.2} M_\odot$, which are, crudely, associated to massive
spheroidal galaxies. We envisage disk (and irregular) galaxies as
associated primarily to halos virializing at $z_{\rm vir} \lsim
1.5$, which have incorporated, through merging processes, a large
fraction of halos less massive than $10^{11.2}\,M_\odot$ virializing
at earlier times, which may become the bulges of late type galaxies.
The model, however, does not follow the formation and evolution of
disk and bulge components of galaxies. For spiral and starburst
galaxies we adopt the phenomenological model described by Negrello
et al. (2007). On the other hand, as shown in the following, these
galaxies are essentially non influential for the purposes of the
present paper in the considered frequency range: proto-spheroids
dominate the contributions both to the power spectrum of
fluctuations and to the angular correlation function of detected
sources.

Because of the strong dilution due to their very broad luminosity
function, the contribution of radio sources to the clustering power
spectrum can be safely neglected in the wavelength range considered
here. Their contribution to Poisson fluctuations was computed using
the De Zotti et al. (2005) model.

\section{Halo Model Formalism}\label{sect:hm}

To compare the clustering properties expected from our model with
observational data we adopt the halo model formalism (Cooray \&
Sheth 2002). The power spectrum of the galaxy distribution is
parameterized as the sum of the 1-halo term, that dominates on small
scales and depends on the distribution of galaxies within the same
halo, and the 2-halo term, that dominates on large scales and is
related to correlations among different halos:
\begin{eqnarray}
P_{\rm gal}(k,z)\!\!\!\!\!\!&=&\!\!\!\!\!\!P^{\rm 1h}_{\rm
gal}(k,z)+P^{\rm 2h}_{\rm gal}(k,z),\hfill \\
P^{\rm 1h}_{\rm gal}(k,z)\!\!\!\!\!\!&=&\!\!\!\!\!\!\int_{M}\!\!\!\!dM\frac{dn}
{dM}\frac{\langle{N_{\rm gal}(N_{\rm gal}-1)}\rangle}{\bar{n}^2_{\rm gal}}|u_{\rm gal}(k,M)|^s,\label{eq:1halo}\\
P^{\rm 2h}_{\rm gal}(k,z)\!\!\!\!\!\!&=&\!\!\!\!\!\!P_{\rm
lin}(k,z)\!\! \left[\!\int_M\!\!\!\!\!
dM\!\frac{dn}{dM}\frac{\langle{N_{\rm gal}}\rangle} {\bar{n}_{\rm
gal}}b(M,\!z)u_{\rm gal}(k,\!M)\right]^2\!\!\!,\label{eq:2halo}
\end{eqnarray}
where $dn/dM$ is the halo mass function (Sheth \& Tormen 1999) and
the linear matter power spectrum, $P_{\rm lin}(k,z)$, has been
computed using the CAMB code\footnote{http://camb.info/} (Lewis,
Challinor \& Lasenby 2000). Here, $u_{\rm gal}(k,M)$ denotes the
Fourier transform of the mass density profile of the galaxy
distribution within the dark matter halo, that we assume to be
approximately the same as that of the dark matter, i.e. we take
$u_{\rm gal}(k,M)\simeq u_{\rm dm}(k,M)$.

${\langle{N_{\rm gal}}\rangle}$ is the mean number of galaxies in a
halo of mass $M$, subdivided in ``central'' and ``satellite''
galaxies (${\langle{N_{\rm gal}}\rangle}= {\langle{N_{\rm
cen}}\rangle}+{\langle{N_{\rm sat}}\rangle}$), while $\bar{n}_{\rm
gal}$ is the mean number density of galaxies:
\begin{equation}
\bar{n}_{\rm gal}=\int_{M}dM\frac{dn}{dM}{\langle{N_{\rm gal}}\rangle}.
\end{equation}
We model the HOD using a central-satellite formalism (see, e.g., Zheng et al. 2005): this
assumes that the first galaxy to be hosted by a halo lies at its
center, while any remaining galaxies are classified as satellites
and are distributed in proportion to
the halo mass profile.
Following Tinker \& Wetzel (2010), the mean occupation functions of
central and satellite galaxies are parameterized as:
\begin{eqnarray}
{\langle{N_{\rm cen}}\rangle}\!\!\!\!\!\!&=&\!\!\!\!\!\!\frac{1}{2}
\left[1+{\rm erf}\left({\frac{\log_{10}(M/M_{\rm min})}{\sigma(\log_{10}M)}}\right)\right],\\
{\langle{N_{\rm
sat}}\rangle}\!\!\!\!\!\!&=&\!\!\!\!\!\!\frac{1}{2}\left[1+{\rm
erf}\left({\frac{\log_{10}(M/2M_{\rm
min})}{\sigma(\log_{10}M)}}\right)\right]\left(\frac{M}{M_{\rm
sat}}\right)^{\alpha_{\rm sat}},
\end{eqnarray}
where $M_{\rm min}$, $\alpha_{\rm sat}$, $M_{\rm sat}$, and
$\sigma(\log_{10}M)$ are free parameters assumed to be redshift
independent. In this formalism halos below $M_{\rm min}$ do not
contain galaxies while halos above this threshold contain a central
galaxy plus a number of satellite galaxies with a power-law mass
function with slope $\alpha_{\rm sat}$.

The mean mass density profile of halos of mass $M$ is (Navarro,
Frenk, \& White 1996):
\begin{equation}
\rho(r)=\frac{\rho_{\rm s}}{(r/r_{\rm s})(1+r/r_{\rm s})^{2}},
\end{equation}\label{eq:NFW}
\begin{equation}
M=4\pi\rho_{\rm s}r^3_{\rm s}\left[\log(1+c)-\frac{c}{1+c}\right],
\end{equation}
with $c=r_{\rm vir}/r_{\rm s}$. The normalized Fourier transform of
this profile is:
\begin{eqnarray}
\lefteqn{u_{\rm dm}(k,M)=\frac{4\pi\rho_{\rm s}r^3_{\rm
s}}{M}\left\{\sin(kr_{\rm s})\left[Si([1+c]kr_{\rm s})-Si(kr_{\rm
s})\right]\frac{}{}\right.}\nonumber \\
& +&\!\!\!\!\!\!  \left.\cos(kr_{\rm s})\left[Ci([1+c]kr_{\rm
s})-Ci(kr_{\rm s})\right]-\frac{\sin(ckr_{\rm s})}{(1+c)kr_{\rm
s}}\right\},
\end{eqnarray}
where $Si$ and $Ci$ are the sine and cosine integrals, respectively:
\begin{equation}
Si(x)=\int^x_0\frac{\sin(t)}{t}dt,~~~Ci(x)=-\int^\infty_x\frac{\cos(t)}{t}dt.
\end{equation}
Following Bullock et al. (2001), we approximate the dependence of
the concentration $c$ on $M$ and $z$ as
\begin{equation}
c(M,z)=\frac{9}{1+z}\left(\frac{M}{M_\ast}\right)^{-0.13}
\end{equation}
where $M_\ast(z)$  is the characteristic mass scale at which
$\nu(M,z)=1$; $M_\ast(z=0)\simeq5\times10^{12}\,h^{-1}\,M_\odot$.

In the 1-halo term [eq.~(\ref{eq:1halo})] we set $s=2$, in analogy
with the corresponding term for the dark matter power spectrum, if
${\langle{N_{\rm gal}(N_{\rm gal}-1)}\rangle} > 1$. Otherwise we set
$s=1$ since if the halo contains only one galaxy, it will sit at the
center. Taking into account that ${\langle{N_{\rm
gal}(N_{\rm gal}-1)}\rangle} \simeq 2{\langle{N_{\rm
cen}\rangle\langle N_{\rm sat}}\rangle}+{\langle{N_{\rm
sat}\rangle}}^2$ and that only the galaxies that are not at the
center get factors of $u_{\rm gal}(k,M)\simeq u_{\rm dm}(k,M)$ we
have:
\begin{eqnarray}
\lefteqn{P^{\rm 1h}_{\rm gal}(k,z)=\frac{1}{\bar{n}_{\rm
gal}^2}\int_{M}dM\frac{dn}{dM} \cdot} \nonumber \\
&\cdot&
\left[2{\langle{N_{\rm cen} \rangle\langle N_{\rm
sat}}\rangle}u_{\rm dm}(k,M)+{\langle{N_{\rm sat}\rangle}}^2u^2_{\rm
dm}(k,M)\right],
\end{eqnarray}
\begin{eqnarray}
\lefteqn{P^{\rm 2h}_{\rm gal}(k,z)=P_{\rm lin}(k,z) \cdot} \nonumber
\\
&\cdot& \left[{\int_M  dM\frac{dn}{dM}\frac{\langle{N_{\rm
gal}}\rangle}{\bar{n}_{\rm gal}}b(M,z)u_{\rm dm}(k,M)}\right]^2,
\end{eqnarray}
with $(dn/dM)dM =f(\nu)(\rho_{\rm m}/M)d\nu$,
\begin{equation}
\bar{n}_{\rm gal}=\int_{M}dM\frac{dn}{dM}{\langle{N_{\rm
gal}}\rangle}=\int_{\nu}d\nu{f(\nu)}\left(\frac{\rho_{\rm
m}}{M}\right){\langle{N_{\rm gal}}\rangle},
\end{equation}
and $\rho_{\rm m}/M=3/(4\pi R^3)$. On large scales, where the 2-halo
term dominates, $u_{\rm dm}(k,M)\rightarrow1$ and $P^{\rm 2h}_{\rm
dm}(k,z)\simeq b_{\rm gal}^2 P_{\rm lin}(k,z)$ with:
\begin{equation}
b_{\rm gal}(z)=\int_{\nu}d\nu{f(\nu)}\left(\frac{\rho_{\rm
m}}{M}\right)b(M,z)\frac{\langle{N_{\rm gal}\rangle}}{\bar{n}_{\rm
gal}}.\label{eq:galbias}
\end{equation}
We also define the effective large-scale bias, $b_{\rm eff}(z)$, as
\begin{equation}
b_{\rm eff}(z) = \int_{M}dM\,\frac{dn}{dM}\frac{\langle{N_{\rm gal}}\rangle}{\bar{n}_{\rm gal}}b(M,z),
\label{eq:beff}
\end{equation}
and the effective mass of the halo, $M_{\rm eff}$,
\begin{equation}
M_{\rm eff}(z) = \int_{M} dM\,\frac{dn}{dM} M \frac{\langle{N_{\rm gal}}\rangle}{\bar{n}_{\rm gal}}.
\label{eq:Meff}
\end{equation}

\begin{figure*}
\begin{center}
\includegraphics[scale=0.28]{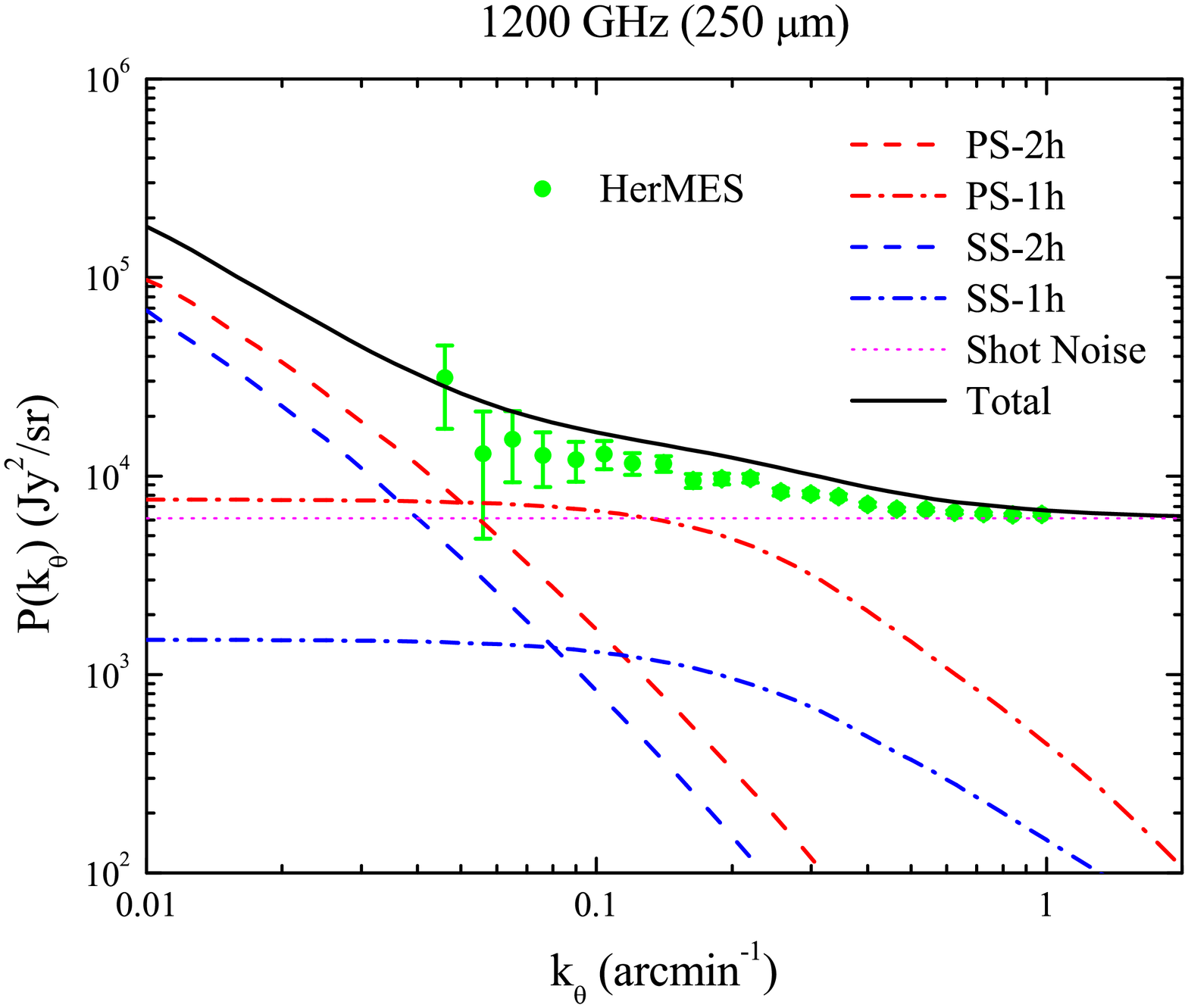}
\includegraphics[scale=0.28]{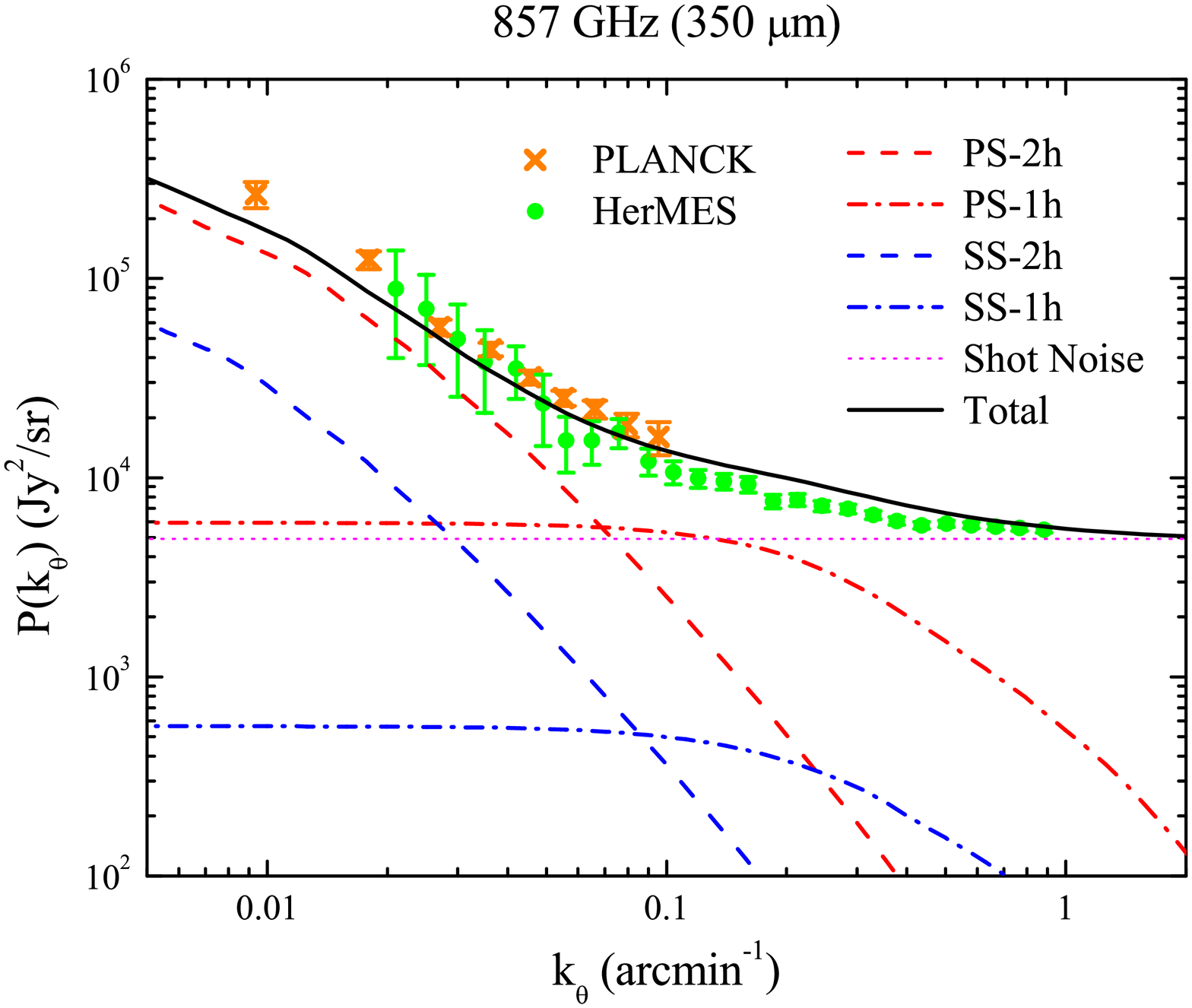}
\includegraphics[scale=0.28]{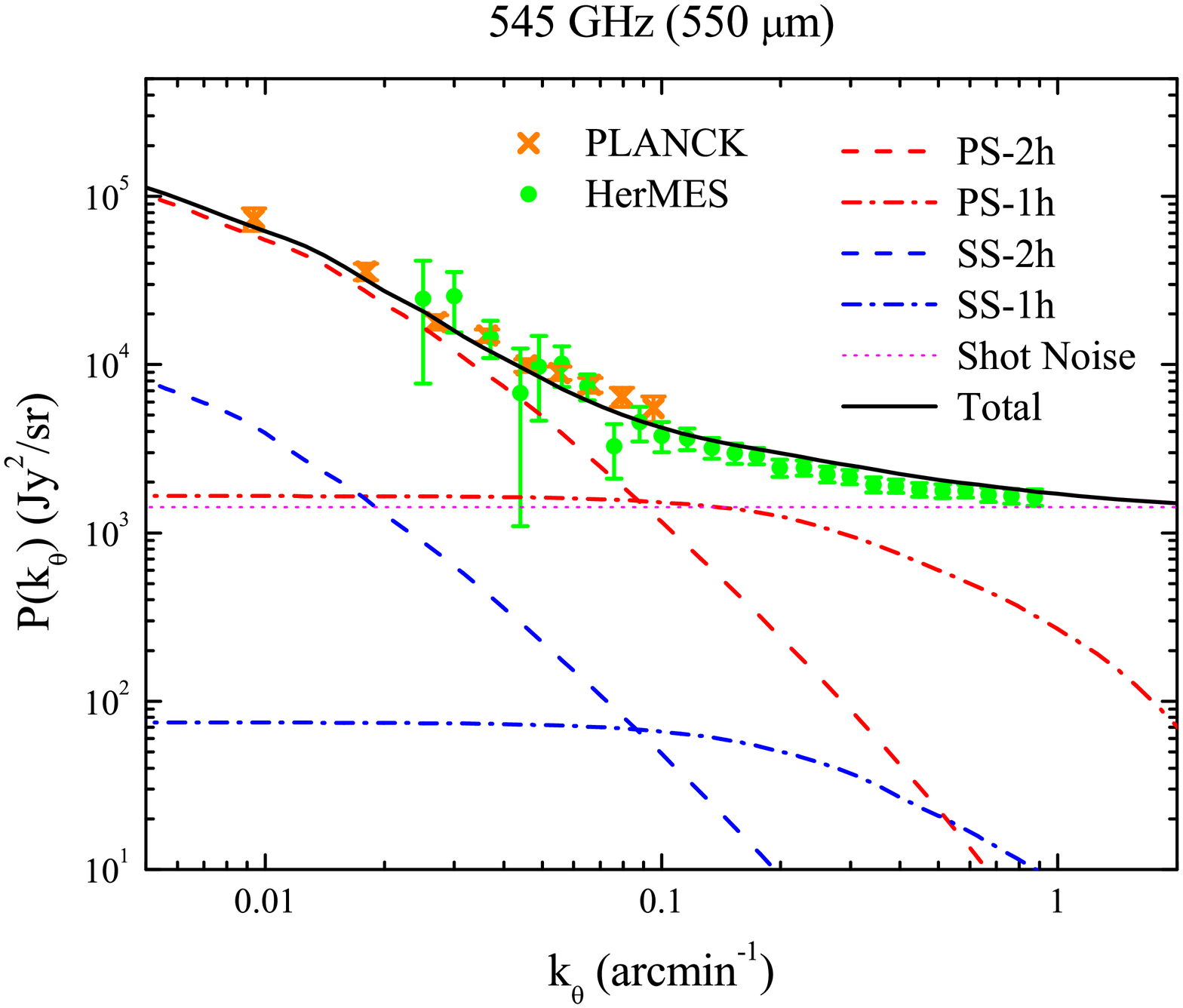}
\includegraphics[scale=0.28]{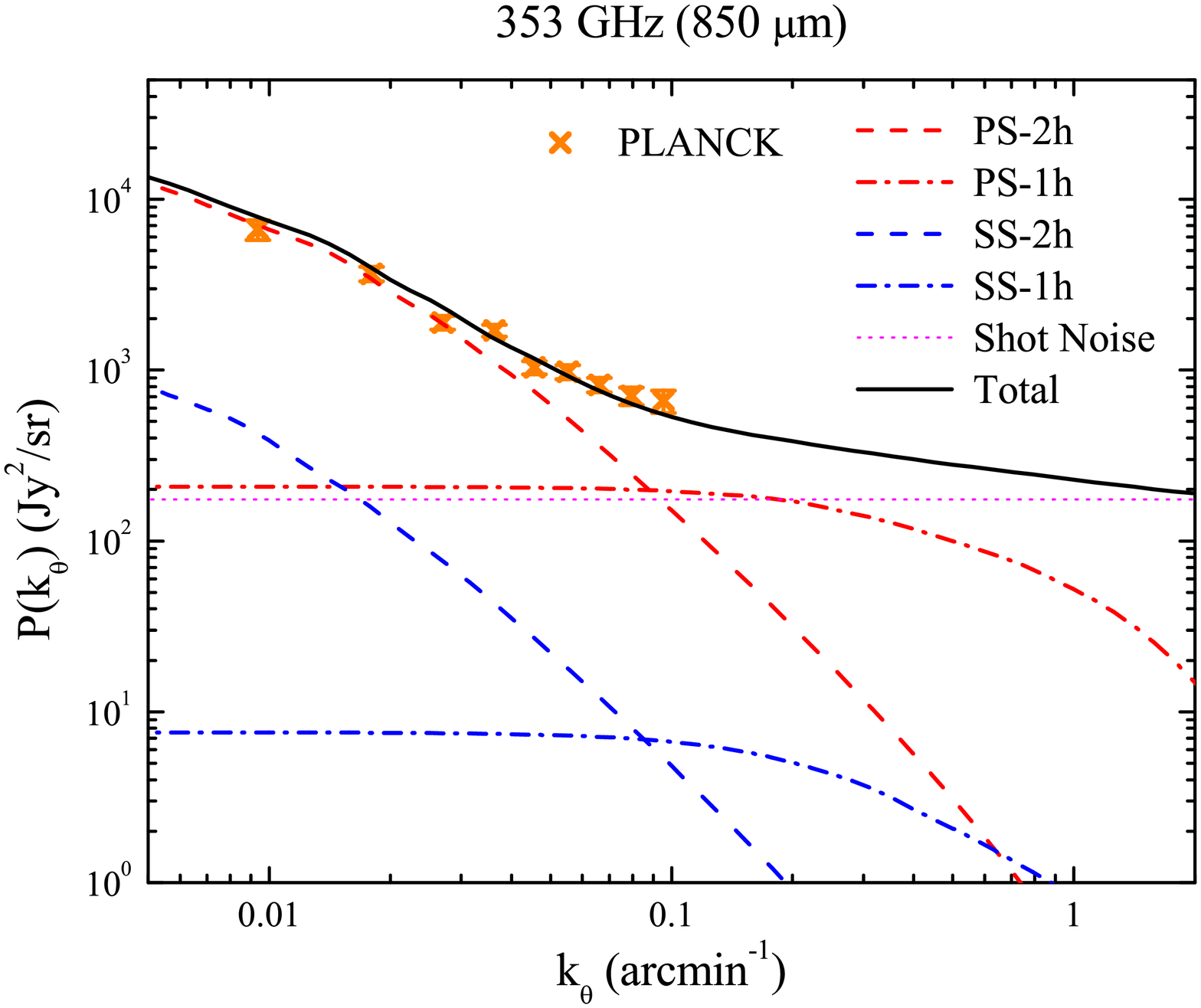}
\caption{CIB angular power spectra $P(k_{\theta})$ at sub-mm
wavelengths. Data from Planck Collaboration (2011) and {\it
Herschel}/HerMES (Amblard et al. 2011). For the {\it
Herschel}/HerMES data at 350 and $500\,\mu$m  we have adopted the
values corrected by Planck Collaboration (2011). At $250\,\mu$m we
have used the values given by Amblard et al. (2011), that may be
underestimated because of an over-subtraction of the cirrus
contamination and a slight overestimate of the effective beam area.
The conversion from the multipole number $\ell$ used by Planck
Collaboration (2011) and the wavenumber $k\,(\hbox{arcmin}^{-1})$ is
$k=\ell/(2\times 180\times 60)$. The lines show the contributions of
the 1-halo and 2-halo terms for the two populations considered here
[spiral and starburst (SS), and proto-spheroidal (PS) galaxies]. The
magenta horizonal lines denote the shot noise level.
}\label{fig:pk_planck}
\end{center}
\end{figure*}

\begin{figure*}
\begin{center}
\includegraphics[scale=0.20]{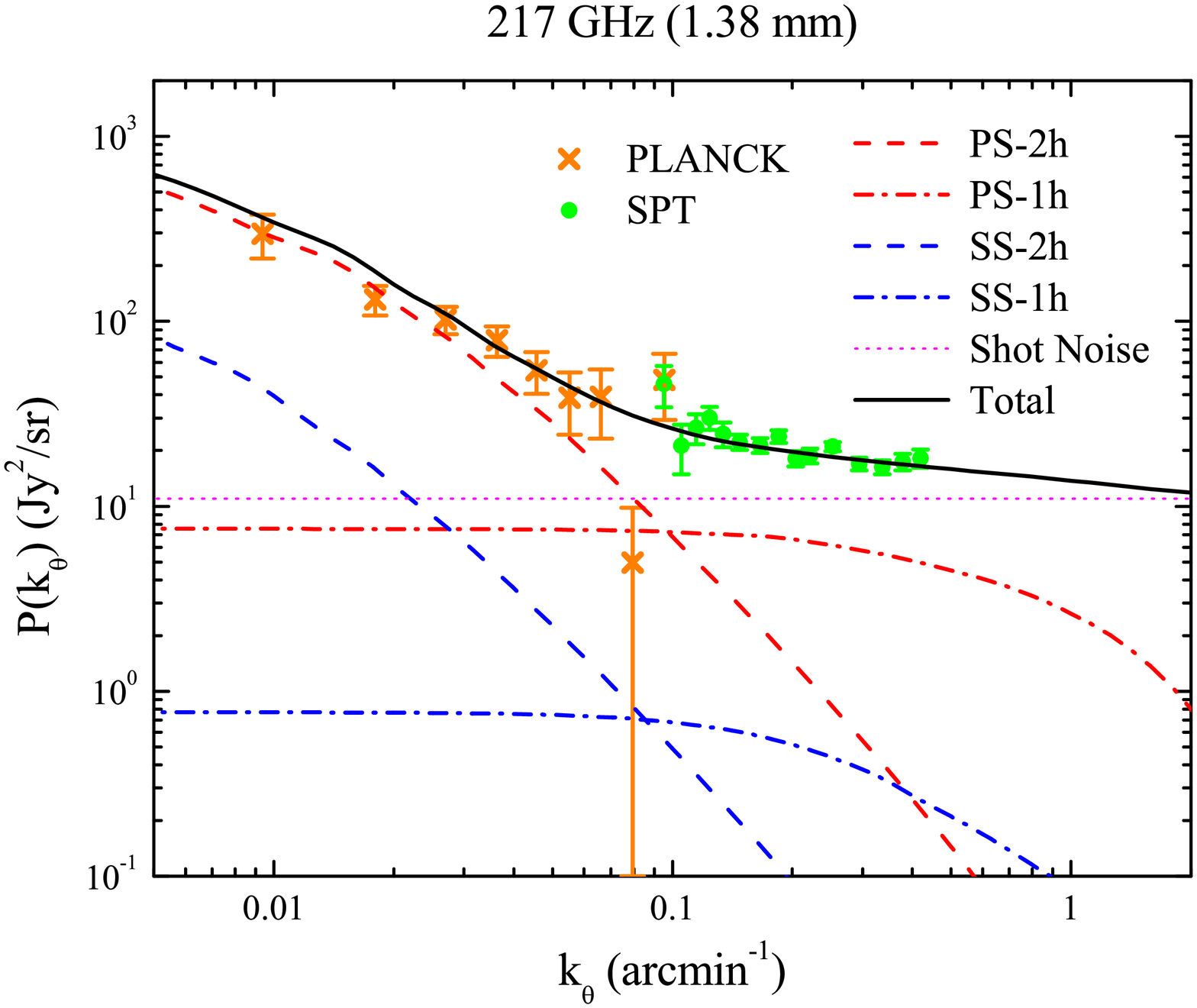}
\includegraphics[scale=0.20]{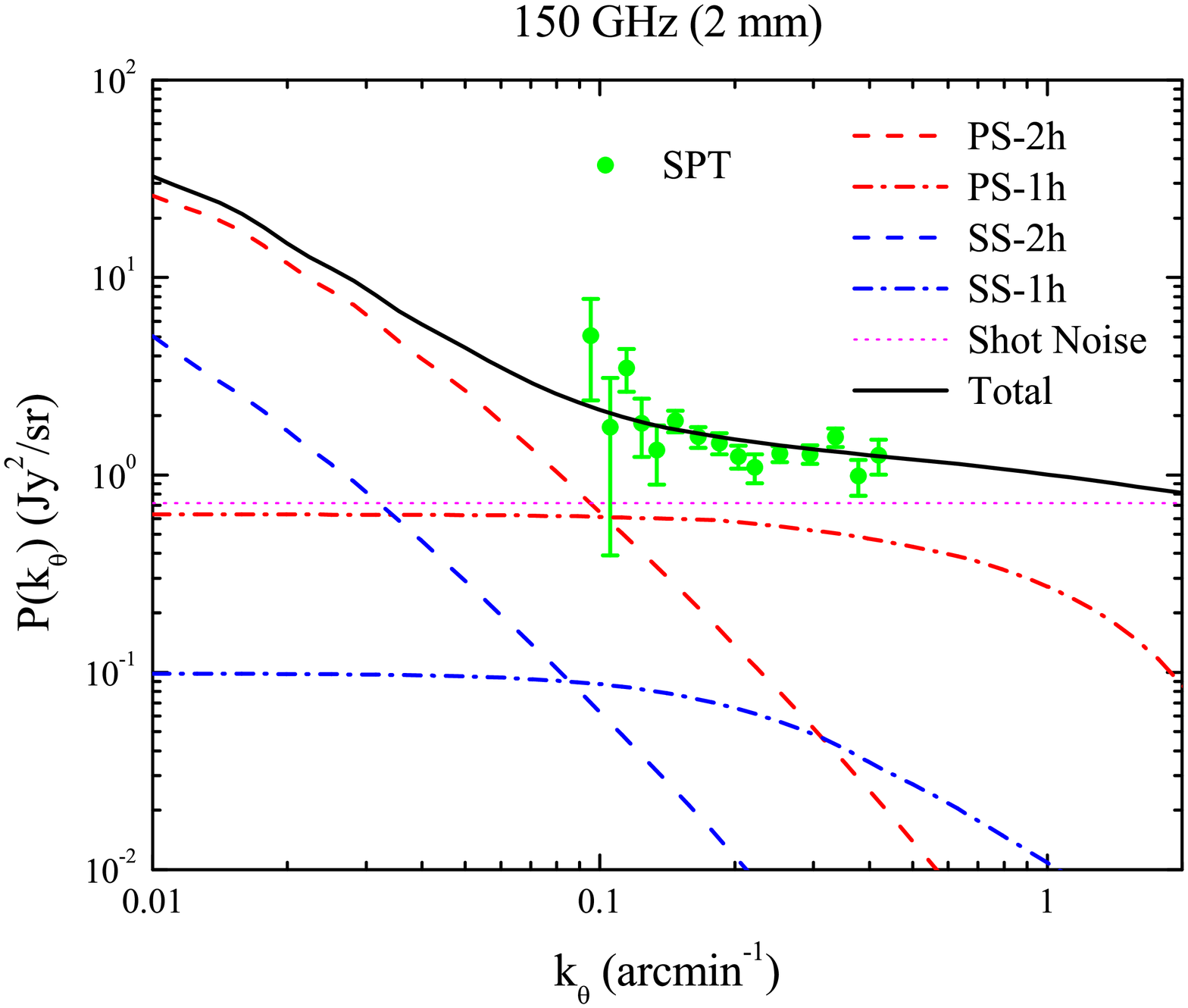}
\includegraphics[scale=0.20]{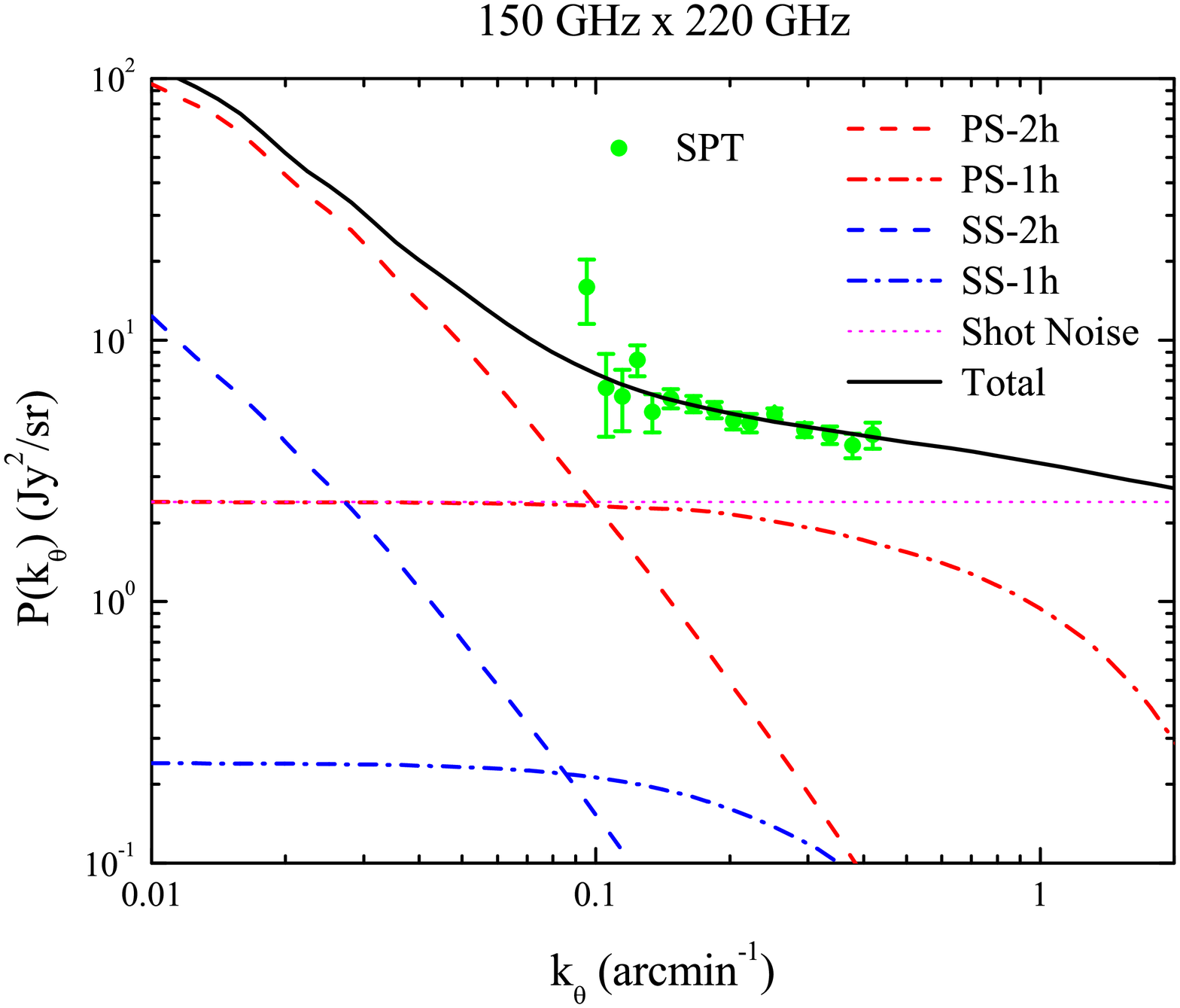}
\caption{CIB angular power spectra $P(k_{\theta})$ at mm wavelengths
and $150\times 220\,$GHz cross spectrum. Data from Planck
Collaboration (2011) and SPT (Hall et al. 2010; Shirokoff et al.
2011). The ACT data (Dunkley et al. 2011; Das et al. 2011) are in
good agreement with the SPT ones and are not plotted to avoid
over-crowding the figure. The lines have the same meaning as in
Fig.~\protect\ref{fig:pk_planck}. }\label{fig:pk_SPT}
\end{center}
\end{figure*}

\section{Angular Power Spectrum of Intensity Fluctuations}\label{sect:ps}

The angular power spectrum $P(k_\theta)$ of intensity fluctuations
due to clustering of sources fainter than some flux density limit
$S_{\rm lim}$ is a projection of the spatial power spectrum of such
sources at different redshifts $z$, $P_{\rm gal}(k,z)$. In the Limber
approximation, valid if the angular scale is not too large (i.e.
$2\pi k_\theta\geq 10$), the relation between $P_{\rm gal}(k,z)$ and
$P(k_\theta)$ is:
\begin{equation}
P(k_\theta)=\!\!\int^{z_{\rm max}}_{z_{\rm min}}\!\!\!\!\!\!\!\!\!\!\!\!dz\,P_{\rm gal}\!\left(\!k=\frac{2\pi
k_\theta+1/2}{\chi(z)},z\right)\left(\frac{dS}{dz}(z)\right)^2\!\!\frac{dz}{dV_{\rm
c}},\label{eq:cl}
\end{equation}
where $dS/dz$ is the redshift distribution of the cumulative flux of
sources with $S\le S_{\rm lim}$
\begin{equation}
{dS\over dz}= \int_0^{ S_{\rm lim}}d\log_{10}(S)\, S\, \phi[L(S,z),z]\,{dV_{\rm c}\over dz},\label{eq:dSdz}
\end{equation}
$\phi(L,z)$ is the epoch-dependent comoving luminosity function per
unit interval of $\log_{10}(L)$, and $dV_{\rm c}$ is the comoving
volume element, $dV_{\rm c}=\chi^2d\chi$, $\chi(z)$ being the
comoving radial distance:
\begin{equation}
\chi(z)=\frac{c}{H_0}\int^z_0\frac{dz'}{\sqrt{\Omega_{\rm m0}(1+z')^3+(1-\Omega_{\rm m0})}}.
\end{equation}
Poisson fluctuations add a white noise contribution to the power
spectrum of fluctuations:
\begin{equation}
P_{\rm shot}=\int^{ S_{\rm lim}}_0 {\frac{dN}{d\log_{10}S}}\,S^2\,{d\log_{10}S},\label{eq:shot}
\end{equation}
with
\begin{equation}
{dN\over d\log_{10}S} = \int\, dz\, \phi[L(S,z),z]\,{dV_{\rm c}\over dz}.
\end{equation}
We have computed the functions $dS/dz$ for each galaxy population
using the the cosmological model specified in \S\,\ref{sec:intro}
and the evolutionary models briefly described in
\S\,\ref{sec:model}. As mentioned in \S\,\ref{sec:model}, the
functions $dS/dz$ for proto-spheroidal galaxies at frequencies $\le
353\,$GHz ($\lambda \ge 850\,\mu$m) have been scaled down by
constant factors to comply with the measurements or the best
estimates of the shot-noise levels. At higher frequencies our model
accurately fits the observed source counts and therefore provides
directly reliable estimates of the shot noise level.

We have chosen to deal with the shot noise and the clustering
contributions to the power spectrum of CIB fluctuations
independently of each other because the former are independent of
the parameters describing the clustering and are strongly
constrained by the available source counts. Moreover, when only
relatively low resolution data are available, as is the case for
{\it Planck}, there is a degeneracy between the shot-noise and the
1-halo clustering term.  As clearly highlighted by Planck
Collaboration (2011), an unsupervised least-square fit of the full
CIB power spectrum measured by {\it Planck}, taking the shot-noise
amplitude as a free parameter, leads to fits of similar quality with
and without a substantial contribution from the 1-halo term. But
fits with a low contribution from the 1-halo term imply shot noise
amplitudes far in excess of those estimated from the source counts.
The higher resolution of {\it Herschel}, SPT and ACT data breaks the
degeneracy at $\nu \ge 600\,$GHz and at $\nu \le 220\,$GHz,
respectively, allowing a direct estimate of the shot-noise
amplitude.

As for the halo model, we have considered two distinct populations,
i.e. proto-spheroidal galaxies and late-type galaxies, both
quiescent and starbursting. Taking into account the constraints on
clustering of late-type galaxies coming from IRAS data (Mann et al.
1995; Hawkins et al. 2001) we find that the contribution of these
sources is always sub-dominant and, correspondingly, their halo
model parameters are very poorly constrained. Moreover the values of
$M_{\rm sat}$ and $\sigma(\log_{10}M)$ are poorly constrained also
for proto-spheroidal galaxies (Planck Collaboration 2011). We have
therefore fixed $M_{\rm sat}=20M_{\rm min}$ and
$\sigma(\log_{10}M)=0.6$ [within the ranges found by Tinker \&
Wetzel (2010) from clustering studies of optical galaxies] for both
populations, and $M_{\rm min, late-type} = 10^{11}\,M_\odot$ and
$\alpha_{\rm sat, late-type} = 1$. We are then left with only 2 free
parameters, i.e. $M_{\rm min}$ and $\alpha_{\rm sat}$, for
proto-spheroidal galaxies.

The angular power spectra of CIB anisotropies at 217, 353, 545, and
857 GHz on the multipole range $200 \le \ell \le 2000$ have been
determined by Planck Collaboration (2011) using {\it Planck} maps of
six regions of low Galactic dust emission with a total area of
$140\,\hbox{deg}^2$. In the same paper, the power spectrum
measurements by Amblard et al. (2011), using {\it Herschel}/SPIRE
data at 250, 350, and $500\,\mu$m and extending down to sub-arcmin
angular scales, i.e. up to $\ell \sim 2\times 10^4$, were
re-analyzed. It was found that Amblard et al. (2011) overestimated
the correction for contamination by Galactic cirrus. Moreover, the
diffuse-emission calibration of SPIRE data was improved using the
more accurate {\it Planck}/HFI calibration. We have used the Amblard
et al. (2011) data as corrected by Planck Collaboration (2011) at
350 and $500\,\mu$m. No correction could be applied at $250\,\mu$m
so that the data points at this wavelength could be underestimated.

Power spectrum measurements at mm wavelengths (around 150 and 220
GHz) have been obtained with the SPT and the ACT (Hall et al. 2010;
Dunkley et al. 2011; Shirokoff et al. 2011; Das et al. 2011). The
subtraction of the other components (CMB, Sunyaev-Zeldovich effect,
radio sources) has been done using the best fit values given in the
papers. Note that the units quoted as $\mu\hbox{K}^2$ are actually
$\mu\hbox{K}^2\times$sr. The conversion factor from these units to
$\hbox{Jy}^2/$sr is $\simeq
[24.8(x^2/\sinh(x/2))]^2/[2\ell(\ell+1)]$. The factor
$[24.8(x^2/\sinh(x/2))]^2$ is $\simeq 1.55\times 10^5$ at 150 GHz
and $\simeq 2.34\times 10^5$ at 217 GHz.


The angular correlation function $w(\theta)$ for a single source
population writes, in terms of the 2D power spectrum $P(k_\theta)$:
\begin{equation}
w(\theta) = 2\pi \int_0^\infty k_\theta P(k_\theta) J_0(2\pi
k_\theta \theta) dk_\theta, \label{eq:wtheta}
\end{equation}
where $J_0$ is the Bessel function of order 0.

Here we have two sub-populations, proto-spheroidal and
spirals$+$starburst galaxies, with different clustering properties.
If their cross-correlations can be ignored the signal for the whole
is given by (Wilman et al. 2003):
\begin{equation}
w_{\rm tot}(\theta)=f^2_{\rm PS}w_{\rm PS}(\theta)+f^2_{\rm SS}w_{\rm
SS}(\theta), \label{eq:wtheta_tot}
\end{equation}
where $f_{\rm PS}$ and $f_{\rm SS}$ are the fractional contributions
of proto-spheroidal and spirals$+$starburst galaxies, respectively,
to the total counts:
\begin{equation}
f_{\rm PS/SS}=\frac{\int dz\mathcal{N}_{\rm PS/SS}(z)}{\int
dz\mathcal{N}_{\rm tot}(z)}~,
\end{equation}
$\mathcal{N}$ being the redshift distribution. Ignoring the
cross-correlations between the two source populations is justified
because of the widely different redshift distributions implied by
the adopted evolutionary model: as mentioned in \S\,\ref{sec:model},
proto-spheroidal galaxies are associated to galactic-size halos
virialized at $z_{\rm vir} \gsim 1.5$ while disk (and
irregular/starburst) galaxies are associated primarily to halos
virializing at $z_{\rm vir} \lsim 1.5$.

The spatial correlation function $\xi(r,z)$ is the Fourier
anti-transform of the 3D power spectrum:
\begin{equation}
\xi(r,z)= {1\over 2 \pi^2} \int_0^\infty k^2\, P_{\rm
gal}(k)\,\left(\sin(kr)\over kr \right)\, dk .
\end{equation}
The clustering radius $r_0(z)$ is defined by $\xi(r_0,z)=1$.

\begin{figure}
\begin{center}
\includegraphics[scale=0.28]{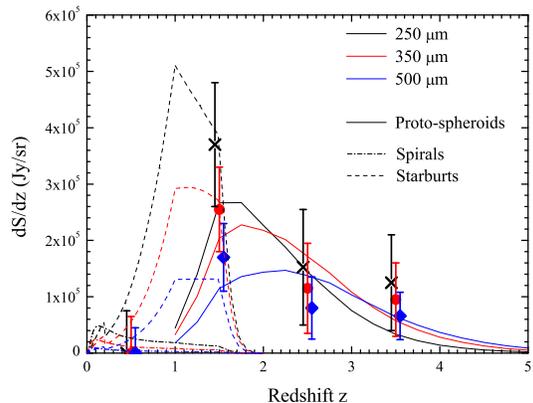}
\caption{Redshift evolution of the galaxy intensities at 250, 350,
and $500\,\mu$m yielded by the model compared with the
observation-based estimates by Amblard et al. (2011). The different
line styles correspond to the different sub-populations, as
specified in the inset. Some caution is needed in interpreting these
data, in view of the problems pointed out by Planck Collaboration
(2011; see text).}\label{fig:dSdz}
\end{center}
\end{figure}


\begin{figure*}
\begin{center}
\includegraphics[scale=0.2]{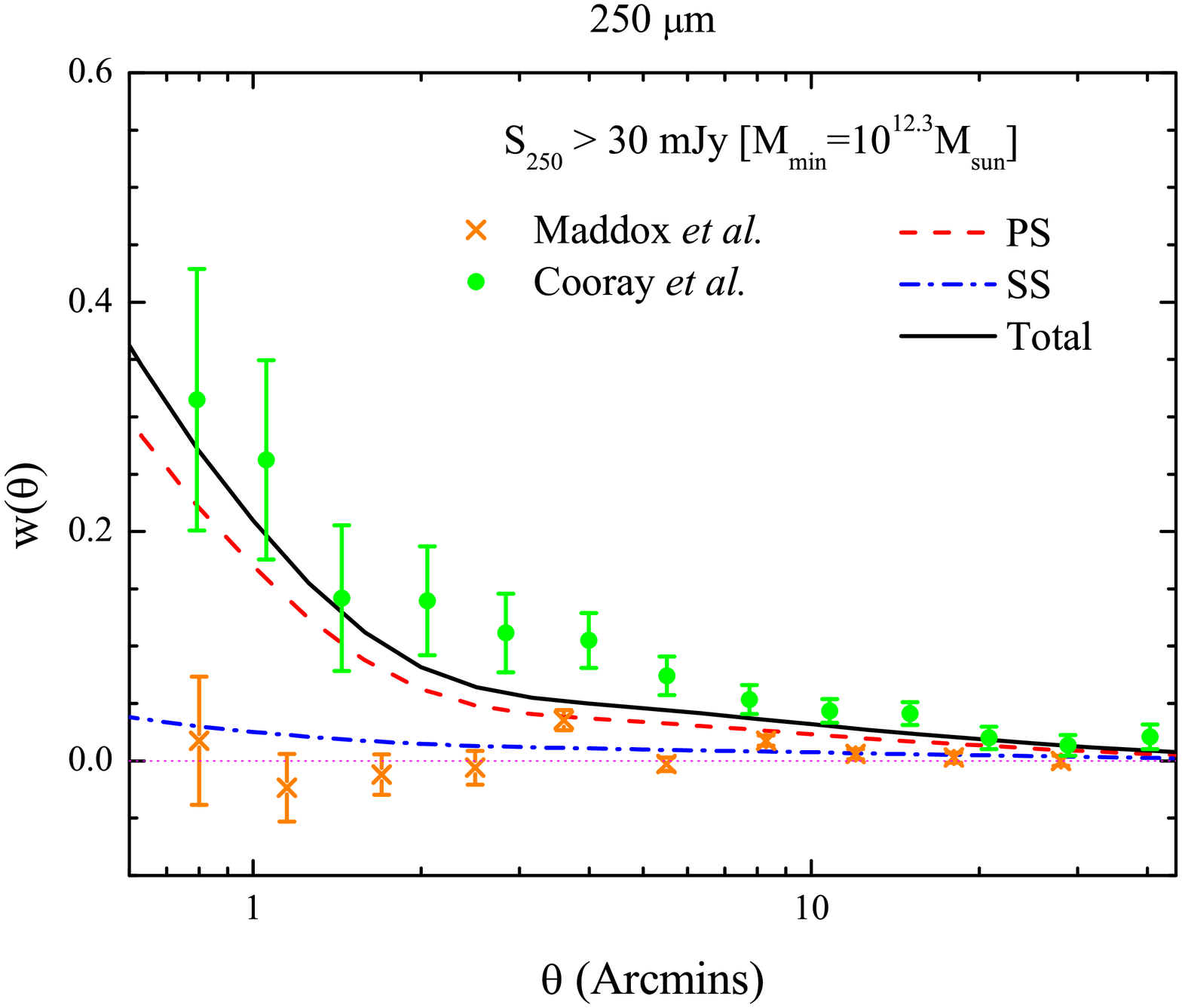}
\includegraphics[scale=0.2]{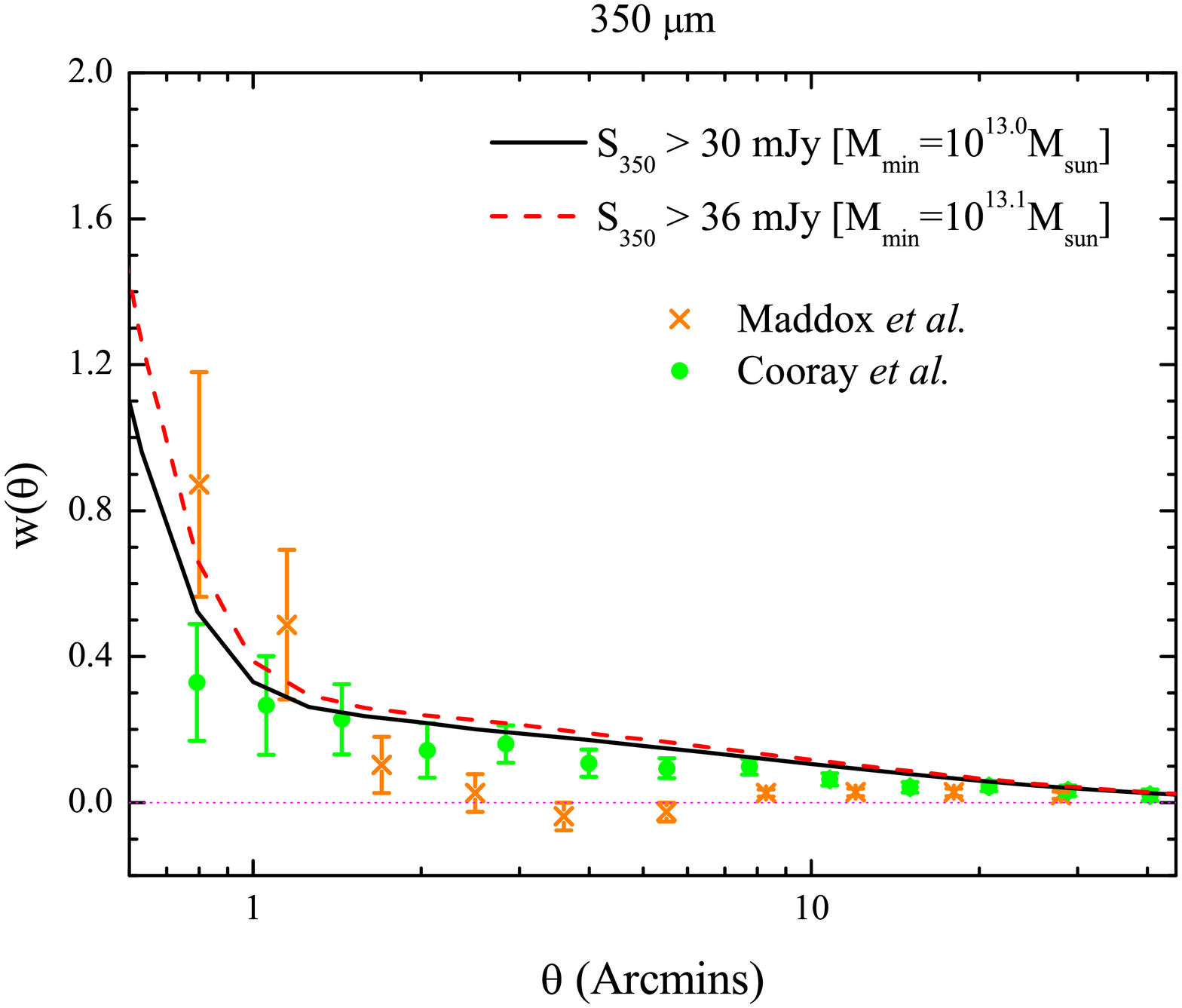}
\includegraphics[scale=0.2]{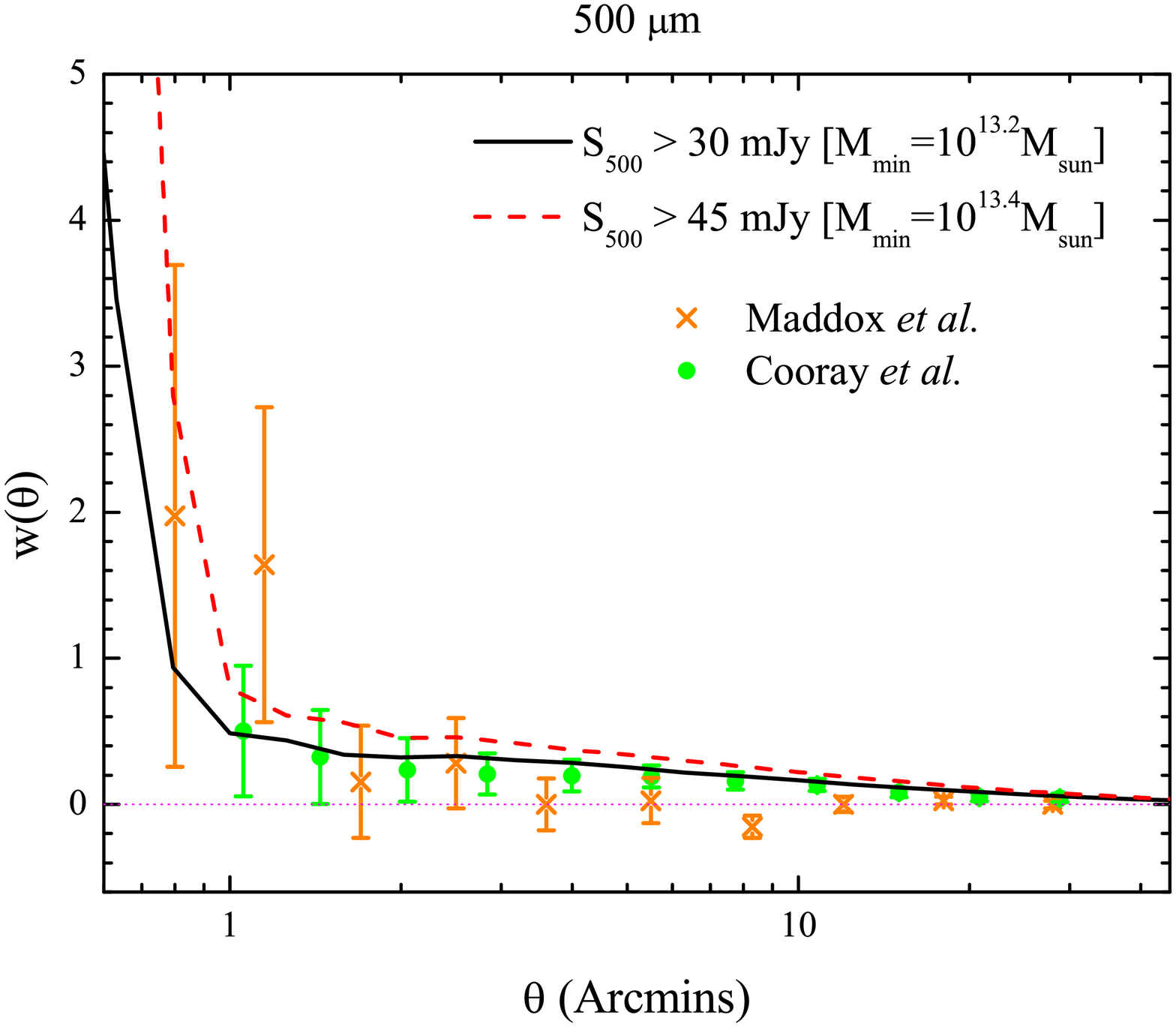}
\caption{Angular correlation function of sub-mm galaxies. Data from
Cooray et al. (2010) and Maddox et al. (2010). In the left-hand
panel the dashed and dash-dotted lines show the contributions of PS
and SS populations, while the solid line denotes the total angular
correlation function. In other two panels, the dashed and solid
lines show the global model correlation functions for the limiting
flux densities adopted in the analyses by Maddox et al. and Cooray
et al., respectively. The horizontal dotted line shows the zero
level. \label{fig:wtheta}}
\end{center}
\end{figure*}

\section{Results}\label{sect:results}

Figures~\ref{fig:pk_planck} and \ref{fig:pk_SPT} compare the best
fit model power spectrum with {\it Planck}, {\it Herschel}, and SPT
data in the wavelength range $250\,\mu$m$-2\,$mm. The agreement is
generally good except at $250\,\mu$m where the model is consistently
above the data points by Amblard et al. (2011) which, however, could
be underestimated (see \S\,\ref{sect:ps}). As mentioned above, we
have only two free parameters, i.e. the minimum mass and the
power-law index of the mean occupation function of satellite
galaxies of proto-spheroidal galaxies. The constraints we obtain are
$\log(M_{\rm min}/M_\odot) = 12.24 \pm 0.06$ and $\alpha_{\rm sat} =
1.81 \pm 0.04$ ($1\,\sigma$). The nominal errors on each parameter
have been computed marginalizing on the other and correspond to
$\Delta \chi^2=1$. We caution that the true uncertainties are likely
substantially higher than the nominal values, both because the model
relies on simplifying assumptions that may make it too rigid and
because of possible systematics affecting the data.

The fact that the same values of these parameters account for the
clustering data from 2 mm to $250\,\mu$m confirms the conclusion by
Planck Collaboration (2011) that CIB fluctuations over this
wavelength range are dominated by a single sub-population of dusty
galaxies. According to our model, this sub-population is made of
proto-spheroidal galaxies making most of their stars at $z>1$. We
find that only at $250\,\mu$m other dusty galaxy populations, normal
disk and starburst galaxies, make a significant, but still
sub-dominant contribution to the clustering power spectrum. As shown
by Lapi et al. (2011), according to our model,  proto-spheroidal
galaxies also account for the bulk of the CIB intensity in this
wavelength range, consistent with the finding by Planck
Collaboration (2011) that the CIB anisotropies have the same
frequency spectrum as the CIB intensity.

Our estimate of the minimum mass is higher than, but consistent,
within the errors, with those found by Amblard et al. (2011)
considering a single galaxy population and 5 free parameters per
frequency (but one of the parameters is unconstrained by the data
within the prior range): $\log(M_{\rm min}/M_\odot) =
11.1^{+1.0}_{-0.6}$ at $250\,\mu$m, $\log(M_{\rm min}/M_\odot) =
11.5^{+0.7}_{-0.2}$ at $350\,\mu$m, and $\log(M_{\rm min}/M_\odot) =
11.8^{+0.4}_{-0.3}$ at $500\,\mu$m.  Our value of $\alpha_{\rm sat}$
is also consistent with those by Amblard et al.: $\alpha_{\rm sat} =
1.6^{+0.1}_{-0.2}$ at $250\,\mu$m, $\alpha_{\rm sat} =
1.8^{+0.1}_{-0.7}$ at $350\,\mu$m and $500\,\mu$m. In Planck
Collaboration (2011) two or three free parameters {\it per
frequency} were used; the derived minimum masses are in the range
$\log(M_{\rm min}/M_\odot) = 11.8-12.5$.

There is however an interesting difference with Planck Collaboration
(2011), due to the different redshift distributions of sources. The
crossover between the 1-halo and the 2-halo term occurs, according
to the model by Planck Collaboration (2011), at multipole numbers
ranging from $\ell \simeq 800$ at 857 GHz ($350\,\mu$m) to $\ell
\simeq 1200$ at 217 GHz ($1.38\,$mm) corresponding to angular scales
ranging from $\theta \simeq 180\times 60/\ell \simeq 13.5'$ at 857
GHz to $9'$ at 217 GHz. According to the B\'ethermin et al. (2011)
model used in that paper, the contribution to the CIB intensity at
857 GHz peaks at $z\simeq 1$ where the angular scale of $13.5'$
corresponds to a physical linear scale $L\simeq 6.5\,$Mpc; at 217
GHz the bulk of the CIB contribution comes from $z>2$ where an
angular scale of $9'$ corresponds to a physical linear scale
$L\simeq 4.5\,$Mpc. The non-linear masses corresponding to an
overdensity $\Delta_c=1.68$ on these scales are $M_{\rm nl}(z=1;
L=6.5\hbox{Mpc})\simeq 7\times 10^{13}\,M_\odot$ and $M_{\rm
nl}(z=2; L=4.5\hbox{Mpc})\simeq 8\times 10^{13}\,M_\odot$,
respectively. For comparison, the characteristic non-linear masses
computed from $\sigma(M_\ast,z)=1.68$ ($\sigma(M_\ast,z)$ being the
rms overdensity) are $M_\ast(z=1)=2 \times 10^{11}\, M_\odot$ and
$M_\ast(z=2)=7.3 \times 10^{9}\, M_\odot$. This suggests that
structures going non-linear on the considered scales are extremely
rare at the corresponding redshifts. This potential difficulty is
eased in our model because the crossover scales are lower by almost
a factor of 2. As shown by Figs.~\ref{fig:pk_planck} and
\ref{fig:pk_SPT}, the 1-halo/2-halo crossover occurs at $\ell \simeq
1450$ at 857 GHz and at $\ell \simeq 2100$ at 217 GHz, corresponding
to angular scales of $7.4'$ and $5.1'$, respectively.

Our value of $M_{\rm min}$ implies an effective halo mass
[eq.~(\ref{eq:Meff})] at $z\simeq 2$ of proto-spheroidal galaxies,
making up most of the CIB, $M_{\rm eff} \simeq 5\times
10^{12}\,M_\odot$, close to the estimated halo mass of the most
effective star formers in the universe. Tacconi et al. (2008)
estimated their mean comoving density at $z\sim 2$ to be $\sim
2\times 10^{-4}\,\hbox{Mpc}^{-3}$. For the standard $\Lambda$CDM
cosmology this implies that they are hosted by dark matter halos of
$\sim 3.5\times 10^{12}\,M_\odot$ (Dekel et al. 2009).

Figure~\ref{fig:dSdz} compares the flux density coming from
different redshifts, $dS/dz$ [eq.~(\ref{eq:dSdz})], predicted by the
model at the SPIRE wavelengths with the best fit estimates by
Amblard et al. (2011). Planck Collaboration (2011) give (their Table
7) the best fit values of the redshift-independent volume
emissivity, $j_{\rm eff}$, for $z>3.5$ [their eq.~(43)]. The values
of $j_{\rm eff}$ given by our model (52, 175, 265, and
$205\,\hbox{Jy}\,\hbox{sr}^{-1}\,\hbox{Mpc}^{-1}$ at 217, 353, 545,
and 857\,GHz, respectively) are consistent with the best-fit
results.

As for the angular correlation function, $w(\theta)$, of detected
SPIRE galaxies, Cooray et al. (2010) reported measurements of
$w(\theta)$ for sources brighter than 30 mJy at all SPIRE
wavelengths and inferred values of $\log(M_{\rm min}/M_\odot)$
ranging from $12.6^{+0.3}_{-0.6}$ at $250\,\mu$m to
$13.5^{+0.3}_{-1.0}$ at $500\,\mu$m. On the other hand, Maddox et
al. (2010) did not detect a significant clustering for their
$250\,\mu$m selected sample with a flux limit of
$33\,\hbox{mJy}\,\hbox{beam}^{-1}$, but detected strong clustering
at $350\,\mu$m and $500\,\mu$m, albeit with relatively large
uncertainties. Our model entails a relationship between the
far-IR/sub-mm luminosity of proto-spheroidal galaxies (that provide
the dominant contribution to $w(\theta)$, see the left-hand panel of
Fig.~\ref{fig:wtheta}) and the associated halo masses (Lapi et al.
2011). For the flux density limit adopted by Cooray et al. (2010),
30 mJy at all SPIRE wavelengths, the model yields $\log(M_{\rm
min}/M_\odot)\simeq 12.3$, 13,  and 13.2 at 250, 350, and
$500\,\mu$m, respectively, while for the flux density limits of
Maddox et al. (2010; 33, 36, and 45 mJy) we have $\log(M_{\rm
min}/M_\odot)\simeq 12.3$, 13.1, and 13.4. The corresponding
predictions for $w(\theta)$ are compared with the data in
Fig.~\ref{fig:wtheta}. The agreement of the model with the data is
generally good, although the situation at $250\,\mu$m is unclear
since there is a discrepancy between the Cooray et al. (2010) and
the Maddox et al. (2010) results.

\section{Discussion and conclusions}\label{sect:conclusions}

According to the Granato et al. (2004) model, the steep portion of
sub-mm counts is dominated by massive proto-spheroidal galaxies in
the process of forming most of their stars on a timescale varying
with halo mass (shorter for more massive galaxies), but typically of
$\simeq 0.7\,$Gyr, i.e. with a duty cycle of $\simeq 0.2$ at
$z\simeq 2$, where their redshift distribution peaks. As shown
above, this model allows us to reproduce the power spectrum of CIB
fluctuations over a broad frequency range, from $250\,\mu$m to a few
mm, with only 2 free parameters. The model also yields an effective
volume emissivity at different redshifts consistent with
observational estimates. The derived effective halo mass, $M_{\rm
eff} \simeq 5\times 10^{12}\,M_\odot$, is close to that estimated
for the most efficient star-formers at $z\simeq 2$.

The multipole number at which the 1-halo term starts exceeding the
2-halo contribution to the clustering power spectrum ranges from
$\ell=1450$ at 857 GHz to $\ell=2100$ at 217 GHz. These values are
almost a factor of 2 higher (and, consequently, the corresponding
angular scales are almost a factor of 2 lower) than those found by
Planck Collaboration (2011). Since, at the redshifts where the
contribution to the CIB intensity peaks, the corresponding masses
are well above $M_\ast$, this difference translates into a much
larger abundance of the relevant halos.

Alternative models make quite different predictions for the
clustering properties of sub-mm galaxies. A widespread view is that
these objects are powered by major merger events. Two major theories
have been worked out in this general framework. One view is that
sub-mm galaxies are massive objects, seen during a short-duration,
intense, merger-induced burst of star formation (e.g. Narayanan et
al. 2009). Since massive galaxies are rare at high-$z$ and, because
of the short duration of the burst, only a small fraction of them
are in the sub-mm bright phase at a given time, this scenario has
difficulty in reproducing the observed counts. This difficulty may
be overcome assuming an extremely top-heavy initial stellar mass
function that would allow much less massive (hence far more
abundant) galaxies to reach the required luminosities (e.g. Baugh et
al. 2005; Lacey et al. 2010).

The clustering implied by the latter scenario has been investigated
by Almeida et al. (2011) who found, at $z = 2$, a comoving
correlation length of $r_0 = 5.6\pm 0.9\,h^{-1}\,$Mpc for galaxies
with $850\,\mu$m flux densities brighter than 5 mJy or an effective
bias factor $b_{\rm eff}=2.3$; for galaxies with $S_{450\mu{\rm
m}}>5\,$mJy they found $b_{\rm eff}=2.1$. Our model implies
$\log(M_{\rm min}/M_\odot)\simeq 12.4$ for sources with
$S_{450\mu{\rm m}}>5\,$mJy and $\log(M_{\rm min}/M_\odot)\simeq
13.0$ for sources with $S_{850\mu{\rm m}}>5\,$mJy. The corresponding
values of the clustering radius and of the effective bias factor are
$r_0 \simeq 11.2\,h^{-1}\,$Mpc, $b_{\rm eff}=4.3$ at $850\,\mu$m,
and $r_0 \simeq 7.3\,h^{-1}\,$Mpc, $b_{\rm eff}=3.1$ at $450\,\mu$m.
The study by Kim et al. (2011) confirms that the clustering data
require a higher amplitude of the 2-halo term, i.e. more massive
halos than implied by the major mergers plus top-heavy initial
stellar mass function scenario.

Dav\'e et al. (2010) investigated the clustering properties of
rapidly star-forming galaxies at $z\simeq 2$ in the framework of a
very different scenario based on cosmological hydrodynamic
simulations whereby the star formation is not powered by mergers but
by steady gas accretion and cooling that can fuel the star formation
for several Gyrs.  In this scenario typical sub-mm galaxies at $z =
2$ live in massive ($\sim 10^{13}\,M_\odot$) halos and have a duty
cycle $\simeq 50\%$. They are expected to be strongly clustered,
with a clustering radius $r_0\sim 10\,h^{-1}\,$Mpc and a bias factor
of $\sim 6$. These values are well in excess of those following from
our analysis which yields, for the bulk of galaxies at $z\simeq 2$,
$r_0 \simeq 6.9\,h^{-1}\,$Mpc, $b_{\rm eff}\simeq 3$.

These results illustrate the power of accurate measurements of the
CIB power spectrum and of the correlation function of galaxies at
(sub-)millimeter wavelengths to discriminate among competing
evolutionary models for the population of dusty galaxies.

\section*{Acknowledgments}
Thanks are due to G. Lagache for clarifications on the CIB power
spectra derived from {\it Planck} data and to the referee for a
careful reading of the manuscript and useful comments. Our numerical
analysis was performed on the Deepcomp 7000 system of the
Supercomputing Center of Chinese Academy of Sciences. We acknowledge
financial support from ASI (ASI/INAF Agreement I/072/09/0 for the
Planck LFI activity of Phase E2) and MIUR PRIN 2009. MV is supported
by ASI/AAE, PD-INFN 51, PRIN INAF and the FP7 cosmoIGM grants.


\end{document}